\documentclass[10pt, a4paper, onecolumn, oneside, notitlepage]{article}
\usepackage[draft]{hyperref}
\usepackage{amsmath}
\usepackage{graphicx}

\title{Diagnosis of the Wavelength Stability of a Mid-Infrared 
Free-Electron Laser}

\author{Xiaolong Wang, 
Yu Qin, 
Takashi Nakajima\footnote{Electronic address: t-nakajima@iae.kyoto-u.ac.jp}, 
Heishun Zen, Toshiteru Kii, \\
and Hideaki Ohgaki\footnote{Electronic address: ohgaki@iae.kyoto-u.ac.jp}
\vspace{2mm}
\\ \it{Institute of Advanced Energy, Kyoto University,}\\
\it{ Gokasho, Uji, Kyoto 611-0011, Japan} }

\date{}

\begin{document}

\maketitle

\renewcommand{\abstractname}{\vspace{-\baselineskip}}

\abstract{
Wavelength stability of free-electron lasers (FELs) is one of the important 
parameters for various applications. 
In this paper we describe two different methods to diagnose the wavelength 
stability of a mid-infrared (MIR) FEL. 
The first one is based on autocorrelation which is usually used to measure 
the pulse duration, and the second one is based on frequency upconversion 
through sum-frequency mixing (SFM). 
}
\vspace{2mm} 


\renewcommand{\thesection}{\Roman{section}} 
\section{INTRODUCTION}
FELs have been deployed for various applications in the wide (from X-ray 
to microwave \cite{OShea:2001gv}) wavelength ranges, many of which are out of 
reach with conventional lasers. 
In the MIR region, the FELs are extensively utilized in spectroscopic 
applications such as molecular vibrational spectroscopy 
\cite{Zimdars:1993tf, Tully:2006}. In such applications, the spectral 
profile of FELs and its stability is one of the important concerns 
of scientists working in this field. 

\begin{figure}
\centerline{
\includegraphics[width=10cm]{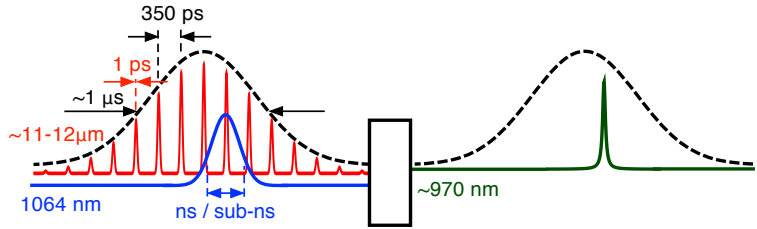}
}
\caption{
Schematic temporal pulse structure of a KU-FEL pulse (red, not in scale), and the temporal shape and timing of an NIR pulse (blue). After the SFM process, we obtain the
SFM pulse (green) at around 970 nm wavelength.}
\label{fig:pulse_structure}
\end{figure}

Unlike any conventional optical laser with gas, liquids, or solid-state 
lasing medium, FELs lase in use of electron beams as a gain medium. 
In Fig. \ref{fig:pulse_structure} we schematically show the temporal 
structure of the emission from Kyoto University free-electron laser (KU-FEL), 
a MIR oscillator-type FEL that runs at 5-13 $\mu$m wavelength \cite{kufel}, 
with an overall macropulse envelope of $\sim$1 $\mu$s duration containing 
a continuous train of much shorter micropulses of $<$1
ps length each, separated from each other by 350 ps intervals.

To determine the spectra and the stability of the FEL in the MIR region, 
conventional approaches involve the use of array detectors or scanning 
monochromators in this spectral region. Employment of such instrument 
is typically costly, and the achievable spectral and temporal resolution 
is somehow limited. In order to obtain comparable or better results 
with more cost-effective instrumentation, we present two different methods 
in this paper, one of which is a method based on the retrieval of spectral 
information from intensity autocorrelation (IAC) and fringe-resolved 
autocorrelation (FRAC) measurements, and the other is based on upconversion 
of MIR pulse spectrum into the NIR region by SFM.

\renewcommand{\figurename}{FIG.}

\section{AUTOCORRELATION-BASED METHOD}

\begin{figure}
\centerline{
\includegraphics[width=10cm]{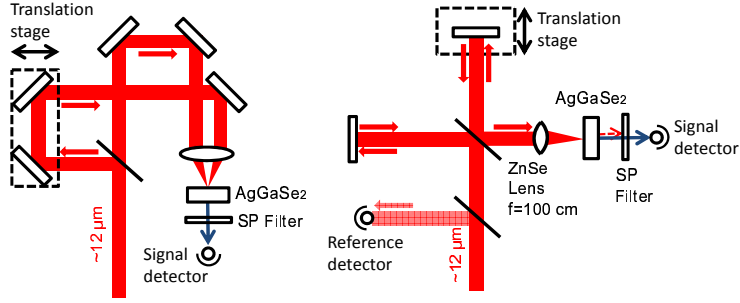}
}
\caption{
Experimental setups for the IAC (left) and FRAC (right) measurements.
}
\label{fig:IAC_FRAC_setup}
\end{figure}

In this section we demonstrate the autocorrelation-based method.
IAC and FRAC have long been used in measurements of ultrashort FEL pulses 
\cite{Diels:1985}.
By investigating the overall envelope of the FRAC signal, we pointed out 
that we are able to derive the wavelength stability of the oscillator-type 
FEL \cite{Qin:2012}. Experimentally, we measure the pulse
duration of the micropulses from KU-FEL by IAC for the first step, whose 
setup is shown in Fig. \ref{fig:IAC_FRAC_setup} together with 
the experimental setup of the FRAC measurement. The autocorrelation fringe 
is measured to be 0.90 ps wide. Under the assumption that the micropulse 
has a Gaussian temporal shape, the micropulse duration is calculated to be 
$\sim$0.64 ps. 
In the FRAC measurement, the FEL beam is split and recombined in
the same way as in a Michelson interferometer. The length of one arm of 
the interferometer is subject to the fine adjustment by scanning a motorized 
translation stage with 2 $\mu$m steps (round trip). As the KU-FEL runs 
at $\sim$12 $\mu$m wavelength, this step length is much smaller than 
the wavelength and is able to resolve the interference 
fringes.The time-delayed pulse pair is focused by a ZnSe lens (f=100 mm) 
onto a piece of nonlinear (AgGaSe$_{2}$) crystal of 1 mm thickness 
to generate the second harmonic (SH). A short-pass filter of 
7.55 $\mu$m cutoff wavelength blocks the fundamental wave and the 
transmitted SH at $\sim$6 $\mu$m wavelength is detected by a joulemeter 
(Gentec. EO, model QE8SP-I-BL-BNC). To avoid the influence of shot-to-shot 
pulse energy fluctuation of the FEL, a small portion of the FEL
pulse is split by a pellicle beam splitter and is recorded by another 
identical joulemeter as a reference signal. We define the autocorrelation 
signal as $S_{\rm FRAC}={S_{\rm Sig}}/{S_{\rm Ref}^{2}}$. 

\begin{figure}
\centerline{
\includegraphics[width=10cm]{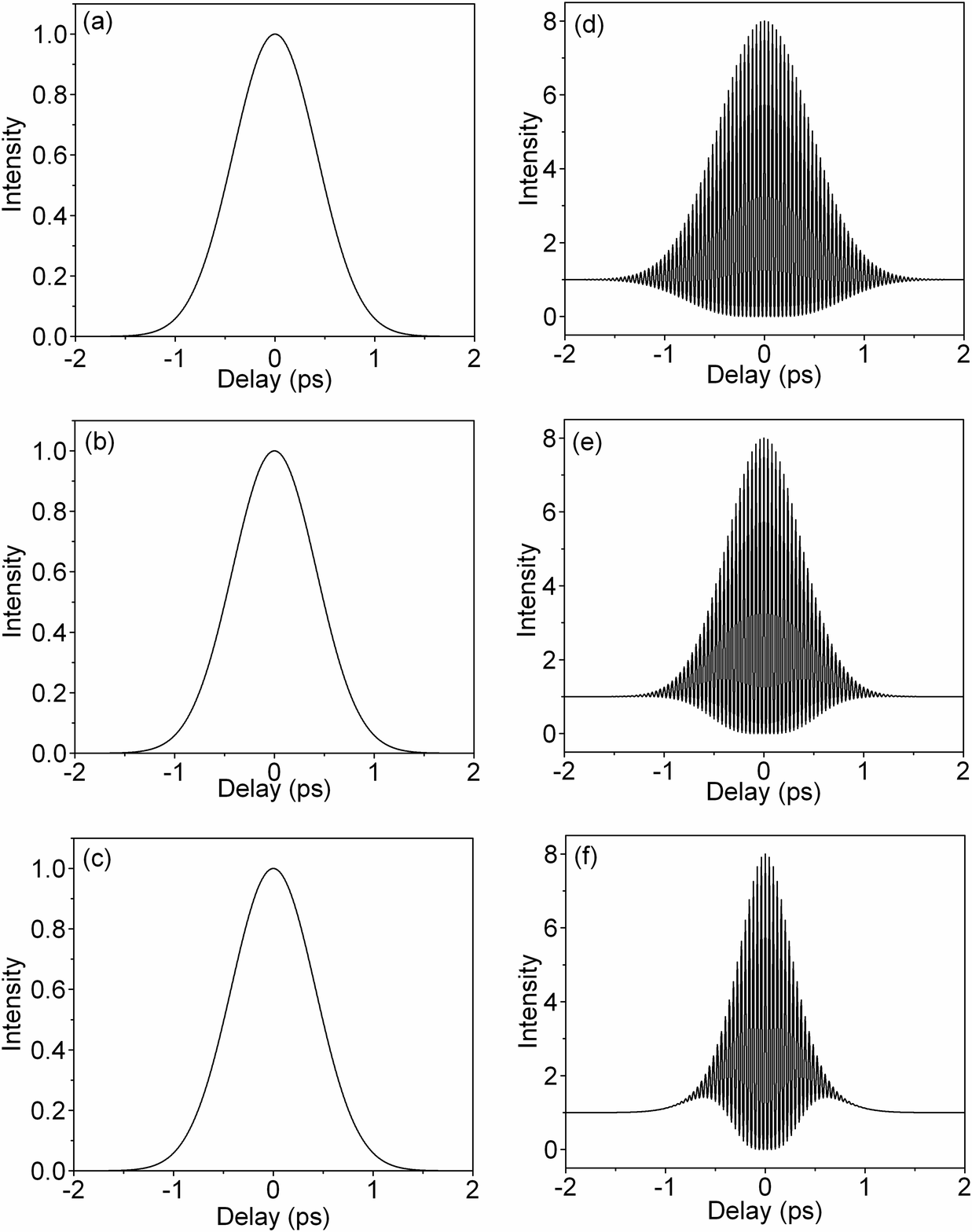}
}
\caption{
Calculated IAC (left column) and FRAC signals (right column) for the 
transform-limited 0.7 ps pulse at 12 $\mu$m with a wavelength stability of 
(a) and (d) $\Delta\lambda/\lambda_{0}$=0\%, (b) and (e) 1\%, and 
(c) and (f) 2\%. 
}
\label{fig:IAC_FRAC_calculation}
\end{figure}

We notice that the deviation of wavelength from the mean FEL wavelength 
changes the autocorrelation signals. The modulation period of the FRAC 
signal is a direct measure of the optical period corresponding 
to the laser wavelength. If various wavelength components exist, they 
induce various modulation periods in their own FRAC signals and the 
summation of such modulated FRAC signals averages out the fringes 
except for those around the center of the signal, where the delay is 
$\tau \sim$ 0. Larger variation of wavelength induces
more narrowing-down on the FRAC signal, so that it is possible to retrieve 
the degree of wavelength fluctuation from the narrowing down of the 
measured FRAC signal width. 
For visualization of this effect, we present  in Fig. 
\ref{fig:IAC_FRAC_calculation} three sets of representative calculated
IAC and FRAC results with the micropulse duration assumed to be 0.7 ps 
and the central wavelength to be 12 $\mu$m. These calculations are done 
with relative wavelength fluctuations ($\Delta\lambda/\lambda_{0}$) of 
0, 1\%, and 2\%, respectively. 

To extract the fluctuation information, we make assumptions that,\\
1. the spectral shape of each FEL micropulse is Gaussian.\\
2. the distribution of the FEL wavelength is Gaussian.\\
3. each micropulse is linearly chirped with unknown chirp rate $\alpha$.

\begin{figure}
\centerline{
\includegraphics[width=10cm]{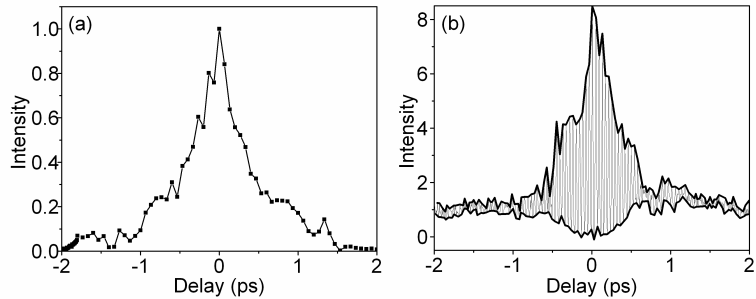}
}
\caption{
Preliminary experimental results for the IAC (left) and FRAC (right) signals. 
The upper and lower envelopes of the FRAC signals are shown by bold lines. 
}
\label{fig:IAC_FRAC_results}
\end{figure}

In Fig. \ref{fig:IAC_FRAC_results} we present preliminary results for the
IAC and FRAC signals obtained at 12 $\mu$m with KU-FEL. 
At present the quality of the signals is rather low, and we must make 
efforts to improve the signal quality.  
But in principle, with the knowledge of micropulse duration obtained 
via IAC measurement, we should be able to fit the upper and lower envelopes 
of the measured FRAC signals to estimate the wavelength stability
of KU-FEL, which is the way we proposed in Ref. \cite{Qin:2012}. Upon trying
the fitting for wavelength stability and chirp rate with the measurement data,
we find that the chirp and the wavelength fluctuation affects the narrowing-down
of the FRAC signal in much the same manner, so that it is difficult to separate
which contribution is how much. Nevertheless, it is
feasible to make a conservative estimation of wavelength fluctuation among
micropulses by assuming the chirp to be negligible. By fitting the obtained envelopes
of the FRAC signal under the no chirp assumption, we find the upper limit
of micropulse wavelength fluctuation to be 1.4\% of the mean central wavelength
of the FEL.

\section{FREQUENCY-UPCONVERSION-BASED METHOD}

In this section we demonstrate the spectroscopic analysis of the FEL spectrum 
by utilizing the method based on upconversion of the 
MIR spectrum into the NIR region, where commercial spectrometers are 
readily available at low cost. The upconversion is realized by the SFM 
process to mix the FEL pulse with another NIR laser pulse. 
Upconversion-based techniques \cite{Shah:1988wl, Heilweil:1989tw} 
have been used in MIR spectroscopy 
\cite{Kubarych:2005io, DeCamp:2005va, Baiz:2011tf, Zhu:2012wq}, and 
upconversion by mixing the THz FEL pulses with a continuous-wave laser has 
recently been reported \cite{Wijnen:2010tb}. In our case, however, 
it is not {\it a priori} obvious whether we can apply this technique 
for the KU-FEL pulses because of the complication of the temporal structure 
of oscillator-type FELs, i.e., the dual pulse structure as shown in 
Fig. \ref{fig:pulse_structure}. 
The train of ultrashort micropulses may damage the nonlinear crystal even if 
the peak intensity of one isolated micropulse is below the damage threshold. 
Also the ultrashort duration of micropulses brings difficulty in 
synchronization between the two lasers. To overcome the effect of timing 
jitter, we use much longer nanosecond or sub-nanosecond NIR laser for the 
SFM process, tolerating the loss of most of the NIR pulse energy 
due to mismatching of the pulse durations. 

\begin{figure}
\centerline{
\includegraphics[width=10cm]{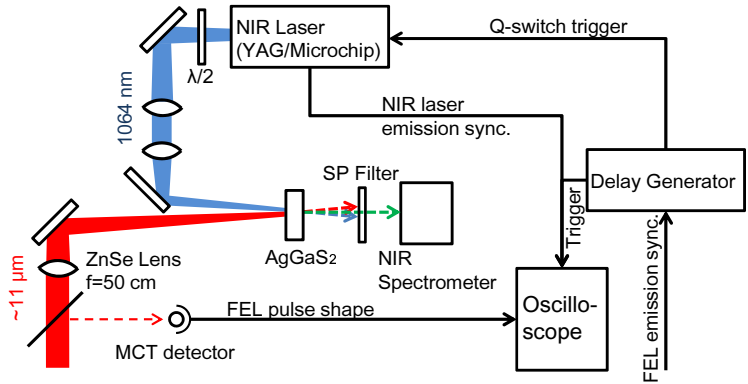}
}
\caption{
Experimental setup for the SFM measurement.
}
\label{fig:experiment_setup}
\end{figure}

The experimental setup for SFM is similar to our recent work presented in 
Ref. \cite{Wang:2012}, and shown in Fig. \ref{fig:experiment_setup}. 
The KU-FEL is synchronized to the Q-switch trigger of an external NIR laser, 
so that both lasers run at 1Hz repetition rate. The delay between 
the FEL and NIR laser pulses is controlled by a delay generator 
(SRI, model DG645), so that the NIR laser pulse coincides with
the FEL pulse on the crystal. A pellicle beam splitter reflects a small 
portion of the FEL pulse onto a mercury cadmium telluride (MCT) 
detector which measures the macropulse shape of the FEL. The pulse shape 
is monitored and recorded on an oscilloscope together with the emission 
synchronization signal from the NIR laser, which gives the information 
of which specific micropulses overlap with the NIR pulse. The recorded 
pulse shape is also used in the offline analysis afterwards. 
We have two choices for the NIR laser source, namely a Q-switched multi-mode 
Nd:YAG laser (LOTIS TII, model LS-2136) and an actively Q-switched 
single-mode microchip laser (Standa, model STANDA-Q1). 
The parameters of the NIR laser employed in the experiment are shown 
in Table \ref{tab:NIR_parameters}. For clarity, all the parameters and 
results presented in this paper involve only the YAG laser as an  
NIR source unless otherwise stated. 

\begin{table}
\centering
\caption{Laser parameters of NIR lasers}\label{tab:NIR_parameters}
\begin{tabular}{ccc}\\ \hline
&YAG laser&microchip laser\\ \hline
emission jitter (ns)&$\sim$50&$<$ 0.35\\ 
pulse duration (ns)&$\sim$20&0.8\\
pulse energy (mJ)&5&0.1\\
diameter on the crystal (mm)&1&0.5\\
intensity at focus (MW/cm$^{2}$)&25&50\\ 
linewidth (nm)&$\sim$0.45&$< 3.7\times10^{-3}$\\
\hline
\end{tabular}
\end{table}

To match the polarization of the FEL and NIR lasers for type I SFM, a 
half-wave plate rotates the linear polarization of the NIR laser to vertical 
direction, identical to that of the FEL. 
The FEL pulse is focused by a ZnSe lens (f = 50 cm), and the NIR pulse 
is also focused by a BK7 lens of the same focal length and reflected 
at right angle to be almost parallel with the FEL, leaving 
0.9$^{\circ}$ angle between the two. In order to avoid damage on the crystal, 
the crystal is placed at off-focus so that the beam diameter on the 
entrance facet is 1 mm. The nonlinear crystal we use is a piece of silver 
thiogallate (AgGaS$_{2}$) of 2 mm thickness, cut at 37$^{\circ}$ angle, 
which guarantees 0.7 $\mu$m wavelength bandwidth and 6.5$^{\circ}$ 
angular bandwidth for the 11 $\mu$m FEL light in the type I SFM process. 
The upconverted pulses at $\sim$970 nm wavelength exit from the opposite 
facet of the crystal and is collected and detected by a commercial VIS-NIR 
spectrometer (OceanOptics, model HR4000CG-UV-NIR). The spectrometer has a 
5 $\mu$m entrance slit and 300 grooves/mm gratings, resulting in 
0.75 nm spectral resolution. 
The 10 $\mu$s exposure (shortest hardware exposure) of the spectrometer 
is triggered by the delay generator, so that the SFM pulses are always 
detected and the stray light from the surroundings is kept minimum, whose
contribution to the background noise is conveniently subtracted 
by the built-in data processing of the spectrometer itself. 
In order to prevent the undesired laser pulses at both $\sim$11 $\mu$m 
and 1064 nm from interfering with the spectral measurement, 
a short-pass filter of 1000 nm cut-off wavelength is placed between 
the exit facet of the crystal and the spectrometer. 

\begin{figure}
\centerline{
\includegraphics[width=10cm]{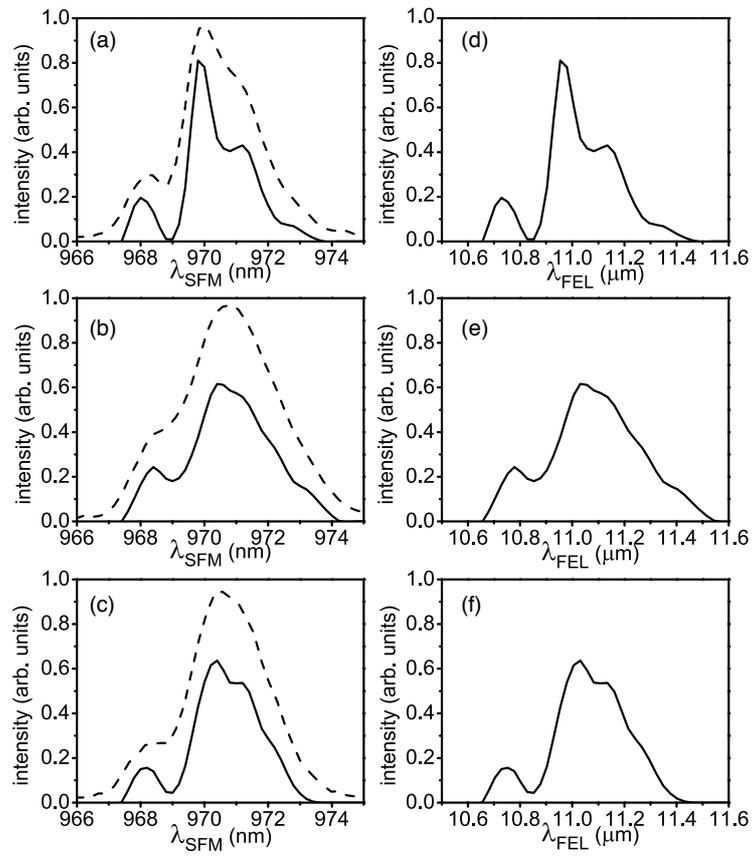}
}
\caption{
(a)-(c) The SFM spectra and (d)-(f) the retrieved FEL spectra 
for three different FEL macropulses.
The dashed and solid curves in (a)-(c) represent the raw and 
deconvolved SFM spectra, respectively. 
}
\label{fig:deconvolution}
\end{figure}

Because of the 0.75 nm spectral resolution of the spectrometer, 
the measured raw spectra of the upconverted 970 nm pulses look wider 
than they actually are. In order to retrieve the actual spectra, 
we measure the spectrum of another narrow-bandwidth laser with the 
spectrometer to obtain its instrumental function, and use it for 
deconvolution of the raw spectra data. After the deconvolution, the 
upconverted spectra in the NIR region is easily transferred back to the 
MIR FEL spectra by the wavelength transformation 
\begin{equation}
\label{wavelength_transformation}
\lambda_{\rm FEL} = {(\lambda_{\rm SFM}^{-1}-\lambda_{\rm NIR}^{-1})}^{-1}.
\end{equation}
Although the 0.75 nm spectral broadening caused by the spectrometer
resolution is reduced by the deconvolution processing, the $\sim$ 0.45 nm
linewidth of the multi-mode YAG laser cannot be removed from the measurement
results, which remains as the main limitation of the spectral resolution of
the SFM-based measurements with conventional multi-mode YAG laser as the NIR light source.
However, the nearly transform-limited microchip laser, which is single-mode and
has much smaller linewidth ($\sim$3.7$\times$10$^{-3}$ nm), can be used in
place of the multi-mode YAG laser and makes this broadening effect negligible.

In Fig. \ref{fig:deconvolution} we present the raw and deconvolved SFM 
spectra (left column), and the retrieved FEL spectra (right column) 
for three different FEL macropulses. 
The NIR pulse emission delay was fixed so that the NIR pulse arrives 
at 1.2 $\mu$s after the intensity maximum of the FEL macropulse for 
these measurements. Although the $\sim$50 ns emission jitter of the YAG
laser makes it difficult to cover exactly the same group of micropulses from one macropulse to
another, and the $\sim$20 ns YAG laser pulse results in the SFM spectrum contributed by
about 57 micropulses which is much more than a single micropulse, it is still acceptable because this 
SFM-based technique is intended to be used in investigating the relatively slow evolution of FEL emission 
conditions in $\mu$s temporal scale, which is more than one order of magnitude larger 
than the temporal uncertainty brought by the jitter of YAG laser or the pulse duration.
The microchip laser which has much smaller emission jitter and pulse duration 
(See Table \ref{tab:NIR_parameters}) can solve
the issues when precise knowledge of the timing is required and when single-micropulse
spectra are under investigation.

The overall spectral resolution of 
the system is limited to $\sim$10 cm$^{-1}$ with the YAG laser, 
and is improved to $\sim$7 cm$^{-1}$ if the transform-limited narrow 
bandwidth microchip laser is used as the NIR source. 
The major limiting factor of the spectral resolution is the resolution 
of the spectrometer itself. By changing to a high-resolution
NIR spectrometer which is available at a reasonable cost, the overall 
resolution can be improved to $\sim$1.2 cm$^{-1}$. 

\begin{figure}
\centerline{
\includegraphics[width=7cm]{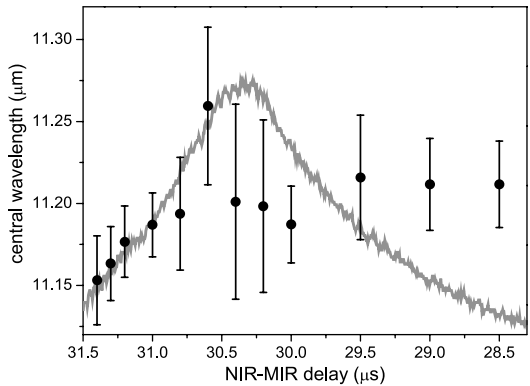}
}
\caption{
Variation of the central wavelength of FEL as a function of delay between 
the YAG and FEL pulses. Data and associated error bars at different delays 
are obtained from the signals by 500 macropulses. The thin grey curve 
in the background shows a typical temporal shape of an FEL macropulse 
to illustrate the timings at different delays. The values of delay 
on the horizontal axis are the actual delay settings in the experiment. 
}
\label{fig:wavelength_delay}
\end{figure}

\begin{figure}
\centerline{
\includegraphics[width=7cm]{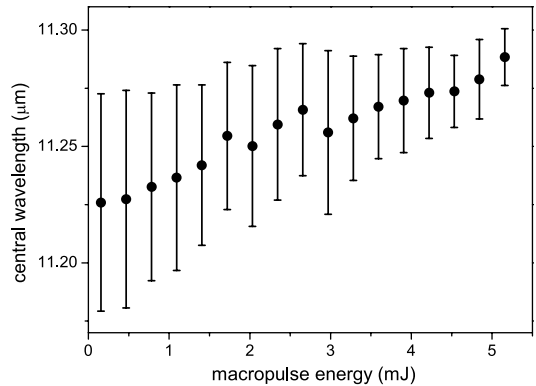}
}
\caption{
Variation of the central wavelength of FEL as a function of macropulse energy. 
Similar to Fig. \ref{fig:wavelength_delay}, data and associated error 
bars at different macropulse energies are obtained from the signals 
by 500 macropulses.
}
\label{fig:wavelength_energy}
\end{figure}

We characterize the FEL wavelength stability by the fluctuation of its 
central wavelength (defined by center-of-mass), and investigate the wavelength
fluctuation by making repeated measurements with different delays between 
the FEL and NIR lasers. 
The results are shown in Fig. \ref{fig:wavelength_delay}. 
We see that even at a fixed timing, the FEL central wavelength has 
large fluctuations, which is in the order of 0.1 $\mu$m, or $\sim$1\% 
of the central wavelength. 
This measured value of wavelength stability is in consistency with 
the evaluation made by the autocorrelation-based method, which estimates
the fluctuation of the central wavelength to be $<$1.4\% of the mean central wavelength.
As the stability value is estimated based on measurements of fixed timing, it
corresponds only to the wavelength fluctuation of the temporally selected $\sim$50
micropulses out of several thousands of micropulses in a single macropulse. 
Fluctuation estimated this way
is reasonably smaller than the FRAC evaluation, which retrieves the fluctuation
over the whole macropulse.

To find the reason of large fluctuation, we now fix the NIR pulse timing 
at the intensity maximum of the FEL macropulse, and plot the measured 
central wavelength with respect to the macropulse energy. 
The results are shown in Fig. \ref{fig:wavelength_energy}. 
Because the fluctuation of FEL macropulse energy results in large error bar, 
we do not find any significant correlation between the central wavelength 
and the macropulse energy. 
Nevertheless, it is clear that the temporal selectivity of the SFM-based 
analysis would provide promising means for further investigation 
to clarify the correlation between the electron beam quality and 
FEL beam quality, etc., which will help understanding the lasing mechanisms 
of FELs.

\section{CONCLUSIONS}
In this paper we have described two different methods to diagnose the 
wavelength stability of KU-FEL, which is an oscillator-type FEL. 
The first method is an autocorrelation-based method, and the FEL 
wavelength fluctuation is extracted through the analysis on the FRAC signals. 
The second method is a frequency-upconversion-based method, which converts
the FEL spectra in MIR region into the NIR region through SFM processes. 
Neither of these methods requires the use of costly instruments 
which are often used for the MIR spectral analysis, such as scanning 
MIR monochromators and array MIR detectors. 
Although we will have to make some efforts to improve the data quality
taken by the first method, we are close to the stage where we are able to 
estimate the wavelength stability of KU-FEL based on the principle 
reported in our previous work \cite{Qin:2012}. 
The employment of the diagnosis methods we have described in this paper 
will help the understanding of the lasing mechanisms. 

\section*{Acknowledgements}
We acknowledge Professor Tetsuo Sakka for the loan of the VIS-NIR 
spectrometer. We also acknowledge Tokyo Instruments Inc. for the loan 
of the Nd:YAG laser.

\end{document}